\documentclass{ws-rv9x6}
\usepackage{subfigure}     
\usepackage{ws-rv-van}     
\makeindex
\newcommand{\beq}{\begin{equation}}
\newcommand{\eeq}{\end{equation}}
\newcommand{\bea}{\begin{eqnarray}}
\newcommand{\eea}{\end{eqnarray}}
\newcommand{\eps}{\varepsilon}
\newcommand{\Vef}{{\cal V}_{\rm eff}}

\begin{document}

\chapter[Microscopic origin of pairing]{Microscopic origin of pairing
}\label{ra_ch1}

\author[Eduard E. Saperstein and Marcello Baldo]{Eduard E. Saperstein and Marcello
Baldo}

\address{Kurchatov Institute, Moscow; saper@mbslab.kiae.ru\\INFN Catania; baldo@ct.infn.it}

\begin{abstract}
A brief review of recent progress in the {\it ab intio} theory of
nuclear pairing is given. Nowdays  several successful solutions of
the {\it ab intio} BCS theory gap equation were published which show
that it is a promising first step in the problem. However, the role
of many-body correlations that go beyond the BCS scheme remains
uncertain and requires further investigations. As an alternative,
the semi-microscopic model  is discussed in which the effective
pairing interaction calculated from the first principles is
supplemented with a small phenomenological addendum containing one
phenomenological parameter universal for all medium and heavy atomic
nuclei.

\end{abstract}

\body

\section{Introduction}

Recently, the fifty years anniversary of Cooper pairing in nuclei
\cite{BMP} took place. However, only in the last few years some
progress has been made in the microscopic theory of nuclear pairing,
first, by the Milan group \cite{milan1,milan2,milan3},  a little
later by Duguet et al. \cite{Dug1,Dug2} and finally by the
Moscow-Catania group \cite{Pankr1,Bald1,Pankr2,Pankr3,Sap1}. In the
first seminal paper of the Milan series, the BCS gap equation for
neutrons with the Argonne v$_{14}$ potential was solved for the
nucleus $^{120}$Sn. The Saxon-Woods Shell-Model basis with the bare
neutron mass $m^*=m$ was used, and the discretization method in a
spherical box was applied to simulate the continuum states
restricted by the limiting energy $E_{\rm max}=600$ MeV. Rather
optimistic result  was obtained for the gap value, $\Delta_{\rm
BCS}=2.2$ MeV. Although it is bigger of the experimental one,
$\Delta_{\rm exp}\simeq 1.3$ MeV, the difference is not so dramatic
and left the hope to achieve a good agreement by developing
corrections to the scheme. In Refs. \cite{milan2,milan3} the basis
was enlarged to $E_{\rm max}=800$ MeV, and, what is more important,
the effective mass $m^*\neq m$ was introduced into the gap equation.
The new basis was calculated within the Skyrme--Hartree--Fock (SHF)
method with the Sly4 force \cite{SLy4}, that makes the effective
mass $m^*(r)$ coordinate dependent and essentially different from
the bare one $m$. E.g., in nuclear matter the Sly4 effective mass is
equal to $m^*=0.7 m$. As it is well known, in the  weak coupling
limit of the BCS theory, the gap is exponentially dependent, i.e.
$\Delta \propto \exp(1/g)$, on the inverse dimensionless pairing
strength $g= m^*\Vef k_{\rm F}/\pi^2$, where $\Vef$ is the effective
pairing interaction. Therefore, a strong suppression of the gap
takes place in the case of $m^*< m$. The value of $\Delta_{\rm
BCS}=0.7$ MeV was obtained in Ref. \cite{milan2} and $\Delta_{\rm
BCS}=1.04$ MeV, in Ref. \cite{milan3}. In both cases, the too small
value of the gap was explained by invoking various many-body
corrections to the BCS approximation. The main correction is due to
the exchange of low-lying surface vibrations (``phonons''),
contributing to the gap around 0.7 MeV \cite{milan2}, so that the
sum  turns out to be $\Delta=1.4$ MeV very close to the experimental
value. In Ref. \cite{milan3}, the contribution of the induced
interaction caused by exchange of the high-lying in-volume
excitations was added either, the sum again is equal to
$\Delta\simeq 1.4$ MeV. Thus, the calculations of Refs.
\cite{milan2,milan3} showed that the effects of $m^*\neq m$ and of
many-body corrections to the BCS theory are necessary  to explain
the difference of ($\Delta_{\rm BCS}-\Delta_{\rm exp}$). In
addition, they are of different sign and partially compensate each
other. Unfortunately, both effects contain large uncertainties. This
point is discussed in Section 3.

In 2008 Duguet and Losinsky \cite{Dug1} made a fresh insight to the
problem by solving the {\it ab initio} BCS gap equation for a lot of
nuclei on the same footing. It should be noticed that the main
difficulty of the direct method to solve the nuclear pairing problem
comes from the rather slow convergence of the sums over intermediate
states $\lambda$ in the gap equation, because of the short-range of
the free $NN$-force.  To avoid the slow convergence, the authors of
Refs. \cite{Dug1,Dug2} used the ``low-k'' force ${\cal V}_{\rm
low-k}$ \cite{Kuo,Kuo-Br} which is in fact very soft. It is defined
in such a way that it describes correctly the $NN$-scattering phase
shifts at momenta $k{<}\Lambda$, where $\Lambda$  is a parameter
corresponding to
 the limiting energy  $\simeq 300\;$MeV.  The force ${\cal V}_{\rm low-k}$
vanishes  for $k{>}\Lambda$, so that in the gap equation one can
restrict the energy range
 to $E_{\max} {\simeq} 300\;$MeV. In addition, a separable version of this force
was constructed that made it possible to calculate neutron and
proton pairing gaps for a lot of nuclei. Usually the low-k force is
found starting from some realistic $NN$-potential ${\cal V}$ with
the help of the Renormalization Group method, and the result does
not practically depend on the particular choice  of ${\cal V}$
\cite{Kuo}. In addition, in Ref. \cite{Dug1} ${\cal V}_{\rm low-k}$
was found starting from the Argonne potential v$_{18}$, that is
different only a little bit from Argonne v$_{14}$, used in Ref.
\cite{milan3}. Finally, in Ref. \cite{Dug1} the same SLy4
self-consistent basis was used as in Ref. \cite{milan3}. Thus, the
inputs of the two calculations look very similar, but the results
turned out to be strongly different. In fact, in Ref. \cite{Dug1}
the value $\Delta_{\rm BCS}\simeq 1.6\;$MeV was obtained for the
same nucleus $^{120}$Sn which is already bigger than the
experimental one by $\simeq 0.3\;$MeV. In Refs. \cite{Bald1,Pankr1}
the reasons of these contradictions were analyzed.  It turned out
that, in fact, these two calculations differ in the way they take
into account the effective mass. It implies that the gap $\Delta$
depends not only on the value of the effective mass at the Fermi
surface, as it follows from the above exponential formula for the
gap, but also on the behavior of the function $m^*(k)$ in a wide
momentum range. However, this quantity is not known sufficiently
well \cite{Bald1}, which makes rather uncertain the predictions of
such calculations.

To avoid such uncertainties, a semi-microscopic model for nuclear
pairing was suggested by the Moscow-Catania group
\cite{Pankr2,Pankr3,Sap1}. It starts from the {\it ab initio} BCS
gap equation with the Argonne force v$_{18}$ treated with the
two-step method. The complete Hilbert space $S$ of the problem is
split into the model subspace $S_0$ of low-energy states and the
complementary one $S'$. The gap equation is solved in the model
space with the effective interaction $\Vef$ which is found  by
projecting out the complementary subspace. A new version of the
local approximation, the so-called Local Potential Approximation
(LPA) \cite{Rep}, is used in the subspace $S'$.  This {\it
ab-initio} term of $\Vef$ is supplemented by a small addendum
proportional to the phenomenological parameter $\gamma$ that should
hopefully embody all corrections to the simplest BCS scheme with
$m^*=m$. Smallness of the correction term is demonstrated in Fig. 1
where a localized ``Fermi average'' form of $\Vef$ is displayed
without ($\gamma=0$) and with ($\gamma=0.06$) the phenomenological
correction. Non-negligible effect of so small change of $\Vef$ to
the gap value is owing  to the above mentioned exponential
enhancement effect. Explicit definition of the functions displayed
in Fig. 1 is given in Section 4 where some results of the
semi-microscopic model are presented.

 \begin{figure}
\centerline {\includegraphics [width=80mm]{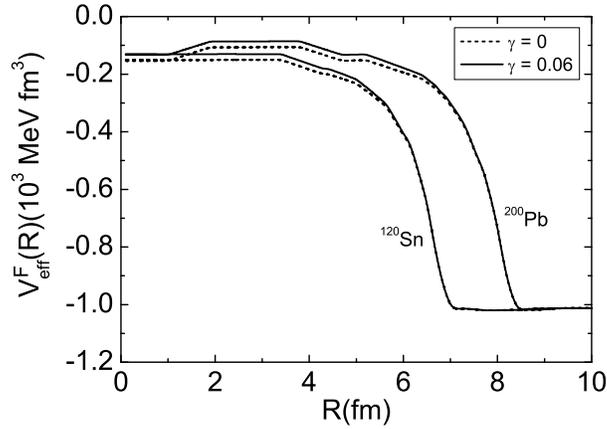}} \vspace{2mm}
\caption{The Fermi average effective pairing interaction ${\cal
V}^{\rm F}_{\rm eff}(R)$ for $^{120}$Sn and $^{200}$Pb nuclei}
\end{figure}

 Fig. 1 demonstrates also the surface nature of nuclear pairing, the
 effective pairing interaction at the surface is ten times stronger
 than inside. This explains the relative  success of the BCS
 approximation. Indeed, at the surface the main corrections to the
 Brueckner-like theory which leads to the BCS scheme are small.

\section{The microscopic BCS equation. LPA approximation}

The general many-body form of the equation for the pairing gap is as
follows \cite{AB},
\beq \Delta_{\tau} =  {\cal U}^{\tau} G_{\tau} G^s_{\tau}
\Delta_{\tau}, \label{del} \eeq where $\tau=(n,p)$ is the isotopic
index, ${\cal U}^{\tau}$ is the $NN$-interaction block irreducible
in the two-particle $\tau$-channel, and
 $G_{\tau}$  ($G^s_{\tau}$) is the one-particle Green function without (with)
 pairing. A symbolic multiplication denotes the integration over
energy and intermediate coordinates and summation over spin
variables as well. The BCS approximation in Eq. (\ref{del}) means,
first, the change of the block ${\cal U}$ of irreducible interaction
diagrams with the free $NN$-potential ${\cal V}$  in Eq.
(\ref{del}), and, second, the use of simple quasi-particle Green
functions $G$ and $G^s$, i.e. those without phonon corrections and
so on. In this case, Eq. (\ref{del}) turns greatly simplified and
can be reduced to the form usual in the Bogolyubov method, \beq
\Delta_{\tau} = - {\cal V}^{\tau} \varkappa_{\tau}\,, \label{delkap}
\eeq where \beq\varkappa_{\tau}=\int \frac {d\eps}{2\pi i}G_{\tau}
G_{\tau}^s\Delta_{\tau}
 \label{defkap}\eeq is the anomalous density matrix
which can be expressed explicitly in terms of the Bogolyubov
functions $u$ and $v$,
\beq \varkappa_{\tau}({\bf r}_1,{\bf r}_2) = \sum_i u_i^{\tau}({\bf
r}_1) v_i^{\tau}({\bf r}_2). \label{kapuv} \eeq Summation in Eq.
(\ref{kapuv}) scans the complete set of Bogolyubov functions with
eigen-energies $E_i>0$.

To overcome the slow convergence problem, a two-step renormalization
method of solving the gap equation in nuclei was used in
Refs.\cite{Pankr2,Pankr3,Sap1}.   The complete Hilbert space of the
pairing problem $S$ is split in the model subspace $S_0$, including
the single-particle states with energies less than a separation
energy $E_0$, and the complementary one, $S'$. The gap equation is
solved in the model space: \beq \Delta_{\tau} = \Vef^{\tau} G_{\tau}
G^s_{\tau} \Delta_{\tau}|_{S_0}, \label{del0} \eeq with the
effective pairing interaction $\Vef^{\tau}$ instead of the block
${\cal V}^{\tau}$ in the BCS version of the original gap equation
(\ref{del}). It obeys the Bethe--Goldstone type equation in the
subsidiary space, \beq \Vef^{\tau} = {\cal V}^{\tau} + {\cal
V}^{\tau} G_{\tau} G_{\tau} \Vef^{\tau}|_{S'}. \label{Vef} \eeq In
this equation, the pairing effects can be neglected provided the
model space is sufficiently large, $E_0\gg \Delta$. That is why we
replaced the Green function $G^s_{\tau}$ for the superfluid system
with its counterpart $G_{\tau}$ for the normal system.  To solve
 Eq. (\ref{Vef}) in non-homogeneous systems
 a new form of the local approximation, the Local
Potential Approximation (LPA), was developed by the Moscow--Catania
group. Originally, it was found for semi-infinite nuclear matter
\cite{Bald0}, then for the slab of nuclear matter (see review
article \cite{Rep}) and, finally, for finite nuclei
\cite{Pankr1,Bald1}. It turned out that, with a very high accuracy,
at each value of the average c.m. coordinate ${\bf R}=({\bf r}_1 +
{\bf r}_2 +{\bf r}_3 +{\bf r}_4)/4$, one can use in Eq. (\ref{Vef})
the formulae  of the infinite system embedded into the constant
potential well $U=U({\bf R})$. This significantly simplifies the
equation for $\Vef$, in comparison with the initial equation for
$\Delta$. As a result, the subspace $S'$ can be chosen as large as
necessary to achieve the convergence. Accuracy of  LPA depends on
the separation energy $E_0$. For finite nuclei, the value of
$E_0{=}40\;$MeV guarantees the accuracy better than 0.01 MeV for the
gap $\Delta$.

Let us notice that the use of the low-k force $V_{\rm low-k}$ could
be also interpreted in terms of the two-step renormalization scheme
of solving the BCS version (${\cal U}{\to} {\cal V}$) of the gap
equation (\ref{del}), with $E_0 {\simeq} 300\;$MeV and with free
nucleon Green functions  $G$ in Eq. (\ref{Vef}) (i.e. $U(R)=0$).
Then, one obtains $\Vef{\to} V_{\rm low-k}$ (see Ref. \cite{Kuo-Br}
where the usual renormalization scheme is used to find $V_{\rm
low-k}$ instead of the Renormalization Group equation).

\section{Corrections to the BCS scheme}
As it was mentioned in the Introduction, there are mainly three
types of corrections to the plain {\it ab initio} BCS gap equation
with bare nucleon mass $m$. The first one is the effect of the
effective mass $m^*\neq m$ considered in Refs.\cite{milan2,milan3}
and Refs.\cite{Dug1,Dug2} as well. The second one is the
contributions of low-lying surface phonons\cite{milan2,milan3} and
the third one, the induced pairing interaction due to high-lying
in-volume excitations.\cite{milan3}

 \begin{figure}
\centerline {\includegraphics [width=80mm]{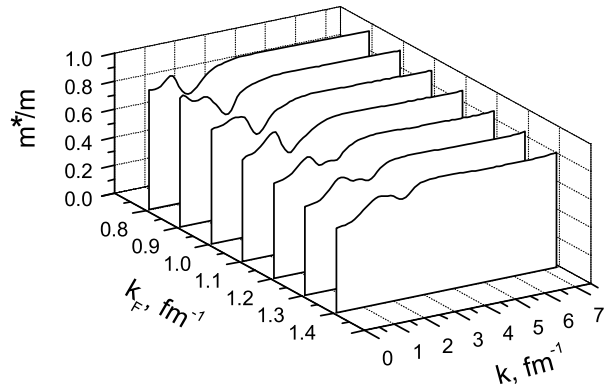}} \vspace{2mm}
\caption{Momentum dependence of the effective mass $m^*(k)$ at
different Fermi momentum values $k_{\rm F}$.}\label{fig:mstk}
\end{figure}

Let us begin from the effective mass. In the
analysis\cite{Pankr1,Bald1} of the difference between the BCS gap
values of Ref.\cite{milan3} and Refs.\cite{Dug1,Dug2}, it was found
that the gap $\Delta$ depends on the behavior of the function
$m^*(k)$ in a wide momentum range. However, this quantity is not
known sufficiently well even in nuclear matter. In Fig. 2 the
effective mass $m^*(k)$  of symmetric nuclear matter at different
Fermi momentum values $k_{\rm F}$ are displayed. They were found
\cite{Bald1} with the Brueckner--Hartree--Fock method which is also,
of course, an approximation. These functions behave rather
non-regular in vicinity of $k_{\rm F}$ and tend to the bare mass
value very slowly. This makes rather doubtful the way to account for
the $m^*$ effect in Refs.\cite{milan2,milan3} and in
Refs.\cite{Dug1,Dug2} as well. In the first case, the effective mass
was taken $k$-independent (equal to $m^*_{\rm SLy4}$) till the
cut-off $k_{\rm max}{\simeq}6\;$fm$^{-1}$. In the second case, the
ansatz was used of $m^*{=}m^*_{\rm SLy4}$ for
$k{<}\Lambda{\simeq}3\;$fm$^{-1}$  and $m^*{=}m$ for $k{>}\Lambda$.

Similar problems appear if one tries to use an explicit form of the
$Z$-factor, which is an additional ingredient of the gap equation,
not only in the combination yielding the complete effective mass
$m^*(k)$ \cite{Bald1}. In this Section, speaking for brevity about
the effective mass we mean in fact both the $k$-mass and $E$-mass.

Corrections to the BCS scheme owing to phonon contributions are of
primary importance.  To our knowledge, the most advanced calculation
of these corrections was carried out in Ref. \cite{milan2}. It
includes as the phonon-exchange term of the block ${\cal U}$ in Eq.
(\ref{del}), the so-called induced interaction, and corrections to
the Green functions $G,G^s$ as well. However, it also possesses some
deficiency connected with omitting the so-called tadpole diagrams.
Up to now, they were consistently taken into account within the
self-consistent Finite Fermi Systems (FFS) theory \cite{Kh-Sap} only
for magic nuclei where pairing is absent. It turned out that their
contribution is usually important and often is of the opposite sign
to the usual diagrams diminishing the total value of the effect
under consideration. This formalism was generalized for superfluid
nuclei in Ref. \cite{Kam_S}, but numerical applications are still
absent.

Let us finally discuss  corrections to the BCS version of Eq.
(\ref{del})  due to the induced interaction from high-lying
particle-hole in-volume excitations. The attempt in Ref.
\cite{milan3} to determine the latter from the SLy4 force together
with the nuclear mean field looks questionable. Indeed, the SLy4
parameters were fitted to the nuclear mass table data mainly related
to  the scalar Landau--Migdal (LM) amplitudes $f,f'$. As to the spin
amplitudes $g,g'$, they remain practically undetermined in the SHF
method. But the contribution of the spin channel to the induced
interaction is not smaller than that of the scalar one
\cite{milan3}. The LM parameters $g,g'$ are well known from the
calculations of nuclear magnetic moments within the FFS theory
\cite{BST} but, as for the Skyrme parameters, only at the Fermi
surface. However, the states distant from the Fermi surface are
important to calculate the induced interaction. The induced
interaction for such states has been determined only in nuclear
matter within the microscopic Brueckner theory \cite{cao}.

\section{The semi-microscopic model for nuclear pairing}

To avoid uncertainties of explicit consideration of corrections to
the BCS scheme discussed above, the semi-microscopic model was
suggested in Refs.\cite{Pankr2,Pankr3,Sap1} In this model, a small
phenomenological addendum to  the effective pairing interaction is
introduceed which embodies approximately all these corrections. The
simplest ansatz for it is as follows: \beq {\cal V}^{\tau}_{\rm
eff}({\bf r}_1,{\bf r}_2,{\bf r}_3,{\bf r}_4) = V^{\rm
BCS}_{\tau,{\rm eff}}({\bf r}_1,{\bf r}_2,{\bf r}_3,{\bf r}_4) +
\gamma^{\tau} C_0 \frac {\rho(r_1)}{\bar{\rho}(0)}
\prod_{i=2}^4\delta ({\bf r}_1 - {\bf r}_i). \label{Vef1} \eeq Here
$\rho(r)$ is the density of nucleons of the kind under
consideration, and $\gamma^{\tau}$ are dimensionless
phenomenological parameters. To avoid any influence of the shell
fluctuations in the value of ${\rho}(0)$, the average central
density ${\bar{\rho}(0)}$ is used in the denominator of the
additional term. It is averaged over the interval of $r{<}2\;$fm.
The first, {\it ab initio}, term in the r.h.s. of Eq. (\ref{Vef1})
is the solution of  Eq. (\ref{Vef}) in the framework of the LPA
method described above, with $m^*{=}m$ in the subspace $S'$.

\begin{figure}
\centerline {\includegraphics [width=80mm]{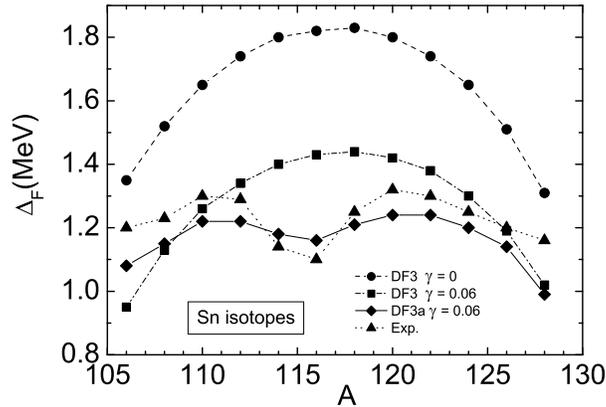}} \vspace{2mm}
\caption{Neutron gap in Sn isotopes }
\end{figure}

We will see below that a rather small value of the phenomenological
parameter $\gamma_n=\gamma_p\simeq 0.06$  is sufficient to produce
 the necessary effect of suppressing theoretical gaps predicted by
the {\it ab initio} calculation.  The smallness of the
phenomenological addendum to the effective
 interaction itself  is demonstrated in Fig. 1 where the
 localized ``Fermi average''
 effective interaction is drawn for $\gamma=0$ and $\gamma=0.06$ values for
 two heavy nuclei. In the mixed coordinate-momentum representation, this
 quantity is defined as
follows: ${\cal V}_{\rm eff}({\bf k}_1,{\bf k}_2,{\bf r}_1,{\bf
r}_2)\to {\cal V}^{\rm F}_{\rm eff}(R=r_1) \prod_{i=2}^4\delta({\bf
r}_1-{\bf r}_i) $, where \beq {\cal V}^{\rm F}_{\rm eff}(R)= \int
d^3t {\cal V}_{\rm eff}(k_1=k_2=k_{\rm F}(R),{\bf R}-{\bf t}/2,{\bf
R}+{\bf t}/2),\eeq with $k_{\rm F}(R)=\sqrt{2m(\mu-U(R))}$, provided
$\mu-U(R)\ge 0$, and $k_{\rm F}(R)=0$ otherwise. Here $\mu$ and
$U(R)$ are the chemical potential and the potential
  well, respectively, of the kind
of nucleons  under consideration. A similar quantity was considered
 before in the nuclear slab to visualize the effective
interaction properties \cite{Rep}.

\begin{figure}
\centerline {\includegraphics [width=100mm]{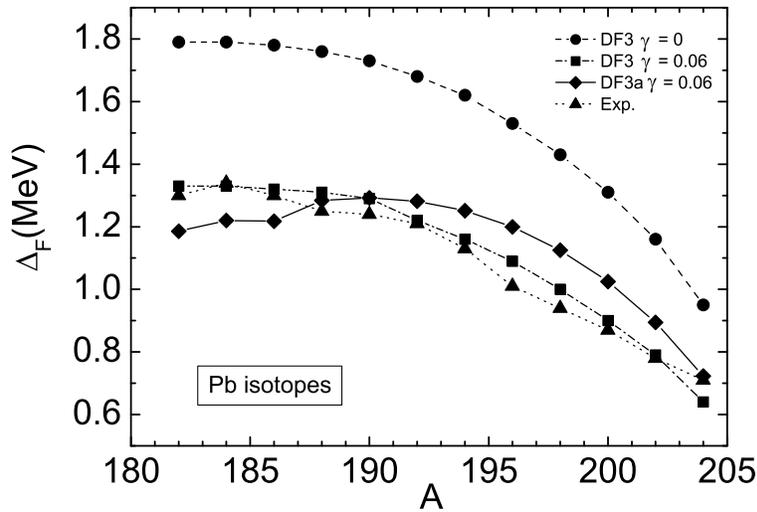}} \vspace{2mm}
\caption{Neutron gap in Pb isotopes. }
\end{figure}

In Ref.\cite{Pankr3}, the above equations were solved in the
self-consistent $\lambda$-basis of the Energy Density Functional
(EDF) of Fayans et al.\cite{Fay1,Fay}. Two sets of the functional
were used, the original one DF3 \cite{Fay} and its modification
DF3-a\cite{Tol-Sap}. In the latter, the spin-orbit and effective
tensor terms of the initial functional were modified. The results
for the pairing gap in three chains of semi-magic nuclei are
displayed in Figs. 2--4.

\begin{figure}
\centerline {\includegraphics [width=100mm]{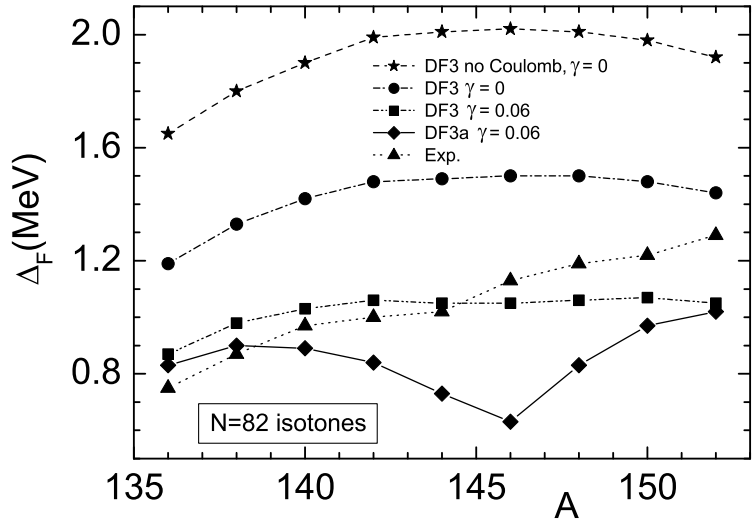}} \vspace{2mm}
\caption{Proton gap for $N=82$ isotones}
\end{figure}

In accordance with the recipe of Ref. \cite{milan3}, we represent
the theoretical gap with the ``Fermi average'' combination \beq
\Delta_{\rm F}=\sum_{\lambda}{(2j{+}1)\Delta_{\lambda
\lambda}}/\sum_{\lambda}(2j{+}1), \label{DelF}\eeq where the
summation is carried out over the states $\lambda$ in the interval
of $|\eps_{\lambda}{-}\mu|{<}3\;$MeV. The ``experimental'' gap is
determined by the symmetric 5-term odd-even mass difference. As it
is argued in Ref.\cite{Pankr3}, the relevance of the mass difference
to the gap has an accuracy of $\simeq(0.1
 \div 0.2)$ MeV. Therefore, it is reasonable to try to achieve
 the agreement of the gap within such limits. It should be noted
 also that the theoretical accuracy of the approach based on
the  ``developed pairing'' approximation \cite{AB} with
 particle number conservation on average only is also about 0.1 MeV
\cite{part-numb}.

Let us begin from neutron pairing and consider first the tin
isotopes, Fig. 2. We see that the BCS gap ($\gamma{=}0$) is
approximately 30\% greater than the experimental one. Switching on
of the phenomenological addendum with $\gamma{=}0.06$ makes
theoretical gap values closer to experiment. However, predictions of
two versions of the functional used are significantly different,
being much better for the DF3-a functional. In particular, the
$A$-dependence of the experimental gap is reproduced  with a
pronounced minimum in the center of the chain. As the analysis in
Ref.\cite{Pankr3} has shown, this strong difference between results
for two functionals is ought to the strong influence to the gap of
the high $j$ intruder state $1h_{11/2}$. Its position depends
essentially on the spin-orbit parameters and is noticeably different
for DF3 and DF3-a functionals. It explains the effect under
discussion.

In the lead chain, see Fig. 3, the overall pattern is quite similar.
Again the BCS gap is approximately 30\% bigger of the experimental
one and again inclusion of the phenomenological term with
$\gamma{=}0.06$ gives a qualitative agreement. Now, the difference
between two functionals is much less. In this case, the agreement is
quite perfect for the DF3 functional and a little worse for the
DF3-a one, but also within limits for the accuracy discussed above.

Let us go to proton pairing, $N{=}82$ chain, see Fig. 4.  In this
case, the Coulomb interaction should be included into the pairing
effective interaction, \beq  \Vef^p=\Vef^n+{\cal V}_{\rm C}.\eeq As
it is argued in Ref.\cite{Pankr3}, the bare Coulomb potential could
be with high accuracy used in this equation. The strong Coulomb
effect in the gap is demonstrated in Fig. 4.  It is also explained
with the exponential dependence of the gap on the pairing
interaction. It should be mentioned that Duguet and co-authors
\cite{Dug2} were the first who inserted the Coulomb interaction
into the pairing force for protons. Only after inclusion of the
Coulomb interaction into $\Vef$, we can use the same value of
$\gamma=0.06$ for protons and neutrons.  As for tin isotopes, the
difference between DF3 and DF3-a results is rather strong, now in
favor of the DF3 functional. This effect is again due to different
positions of the $1h_{11/2}$ level, but now for protons. Overall
agreement with experiment is for protons worse, maybe, because of
closeness of some nuclei to the region of the phase transition to
deformed state.

\section{Conclusions}

 We reviewed briefly the recent progress in the microscopic theory
 of pairing in nuclei involving contributions of the Milan
 group\cite{milan1,milan2,milan3} of Duguet with
 coauthors\cite{Dug1,Dug2} and of Moscow-Catania
 group\cite{Pankr2,Pankr3,Sap1}. It became clear that the plain
{\it ab initio} BCS gap equation with bare mass $m^*=m$
\cite{milan1,Pankr3} is a good starting point for such theory. As
the analysis in Refs.\cite{milan2,milan3} showed,  the effect of
$m^*\neq m$ and that of many-body corrections to the BCS theory are
of different sign and partially compensate each other.
Qualitatively, they can explain the difference between $\Delta_{\rm
BCS}$ and $\Delta_{\rm exp}$  but both effects contain large
uncertainties and hardly can be took into account definitely at the
modern level of nuclear theory.

As an alternative, the semi-microscopic model was suggested in
Refs.\cite{Pankr1,Pankr2,Pankr3} starting from the plain {\it ab
initio} BCS theory with the use of the self-consistent EDF basis
characterized by the bare nucleon mass. The gap equation is recast
in the model space $S_0$, replacing the bare interaction with the
effective pairing interaction ${\cal V}_{\rm eff}$ determined in the
complementary subspace $S'$. The Argonne v$_{18}$ potential was
adopted to find ${\cal V}_{\rm eff}$ along with the LPA
method\cite{Rep}. A small phenomenological term is added to this
effective interaction that contains one parameter, common to
neutrons and protons,  which should embody approximately the
effective mass and other corrections to the pure BCS theory.
Calculations were carried out with two versions of the EDF, the
initial DF3 functional \cite{Fay} and its version DF3-a
\cite{Tol-Sap} with modified spin-orbit and effective tensor terms.
They involve semi-magic lead and tin isotopic chains and the $N=82$
isotonic chain as well. The Coulomb interaction is explicitly
included in the proton gap equation. It was found that  the model
reproduces reasonably well the experimental values of the neutron
and proton gaps for both functionals. However, the results depend
essentially on the single-particle spectrum of the self-consistent
basis used, especially on the position of high $j$-levels. Thus, for
tin isotopes agreement is much better for the DF3-a functional which
reproduces better the position of the ``intruder'' $1h_{11/2}$
neutron level. On the contrary, for $N=82$ isotones agreement for
this functional is worse. In this case, the DF3 functional describes
better the position of the same intruder state but for protons. The
overall disagreement is $\sqrt{\overline{(\Delta_{\rm th}-
\Delta_{\rm exp})^2}}{\simeq}0.13\;$MeV for the DF3 functional and
${\simeq}0.14\;$MeV, for the DF3-a one.

The model under discussion exhibits a week point by including to the
``universal'' phenomenological addendum all corrections to the BCS
scheme. The effective mass correction and the one due to the induced
interaction from  high-lying excitations are mainly in-volume.
Therefore, indeed, they should be universal for medium and heavy
nuclei. On the contrary, the phonon correction is surface and may
vary from one nucleus to another as low-lying phonon characteristics
do. A more consistent scheme should, evidently, include the explicit
consideration of the low-lying phonons, as e.g. in \cite{milan2},
but  taking into account the tadpole diagrams \cite{Kam_S}. In this
case, the phenomenological constant $\gamma$ should, of course,
change. Such extended model is much more complicated but should be
more accurate in reproducing experimental data.

\section*{Acknowledgments}
 We are thankful to U. Lombardo, S. S. Pankratov and M. V. Zverev which
 are coauthors of articles which are the basis of the present review.
 We thank also  G. L. Colo, T. Duguet, V. A. Khodel and S. V. Tolokonnikov for
valuable discussions. The work was partly supported by the DFG and
RFBR Grants Nos.436RUS113/994/0-1 and 09-02-91352NNIO-a, by Grant
NSh-7235.2010.2   of the Russian Ministry for Science and Education,
and by the RFBR grant 11-02-00467-a.

{}

\end{document}